\PassOptionsToPackage{twoside=false}{geometry}
\documentclass[manuscript, screen, nonacm]{acmart}
\usepackage{subcaption} 

\begin{document}

\title{\textsc{Tracer}: A Forensic Framework for Detecting Fraudulent Speedruns from Game Replays}

\author{Jaeung Franciskus Yoo}
\email{franciskus@korea.edu}
\orcid{0009-0004-1136-1693}
\affiliation{
	\institution{Korea University}
	\country{Republic of Korea}
}
\author{Huy Kang Kim}
\email{cenda@korea.ac.kr}
\orcid{0000-0002-0760-8807}
\affiliation{
	\institution{Korea University}
	\country{Republic of Korea}
}

\thanks{\textit{This preprint expands on our Work-in-Progress paper published in the ACM CHI PLAY 2025 Companion Proceedings. The published paper is available at https://doi.org/10.1145/3744736.3749335.}}

\begin{abstract}
Speedrun, a practice of completing a game as quickly as possible, has fostered vibrant communities driven by creativity, competition, and mastery of game mechanics and motor skills. However, this contest also attracts malicious actors as financial incentives come into play. As media and software manipulation techniques advance -- such as spliced footage, modified game software and live stream with staged setups -- forged speedruns have become increasingly difficult to detect. Volunteer-driven communities invest significant effort to verify submissions, yet the process remains slow, inconsistent, and reliant on informal expertise. In high-profile cases, fraudulent runs have gone undetected for years, allowing perpetrators to gain fame and financial benefits through monetised viewership, sponsorships, donations, and community bounties. To address this gap, we propose \textsc{Tracer}, \textit{Tamper Recognition via Analysis of Continuity and Events in game Runs}, a modular framework for identifying artefacts of manipulation in speedrun submissions. \textsc{Tracer} provides structured guidelines across audiovisual, physical, and cyberspace dimensions, systematically documenting dispersed in-game knowledge and previously reported fraudulent cases to enhance verification efficiency.
\end{abstract}

\keywords{Speedrun, ludic metagame, video game, gameplay manipulation, gameplay forensic, collective intelligence based detection}

\maketitle
\nocite{yoo2025tracer}

\section{Introduction}
Challenging oneself and others is a part of human nature: gamers have been showing their enthusiasm in multiple ways such as e-sports tournaments and high score competitions. 
Gamers also often challenge themselves by imposing additional constraints beyond the game's intended mechanics -- a hallmark of what Hemmingsen terms the \textit{ludic metagame}, in which players ``play a game on top of another game''~\cite{refMHLudicMetagame}. 
This form of metagaming includes actions that reinterpret or subvert developers' intentions -- reflecting not only creativity and discipline but, at times, absurd levels of determination~\cite{refMHCodeIsLaw}.

\subsection{The Rise of Ludic Metagame}
\begin{figure*}
	\centering
	\begin{subfigure}[b]{0.45\textwidth}
		\includegraphics[width=\linewidth]{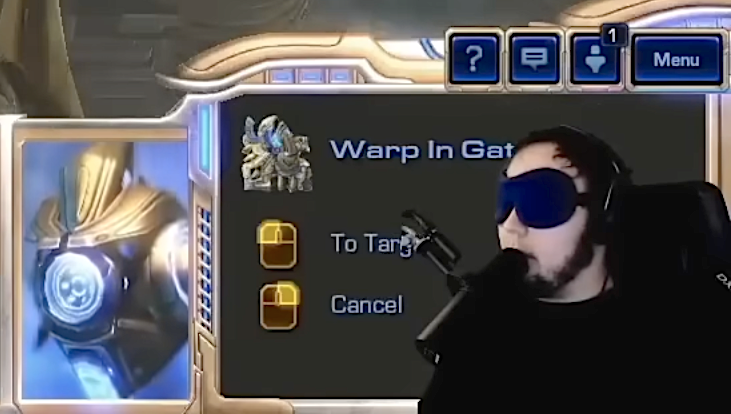}
		\caption{GiantGrantGames}
	\end{subfigure}
	\hfill
	\begin{subfigure}[b]{0.45\textwidth}
		\includegraphics[width=\linewidth]{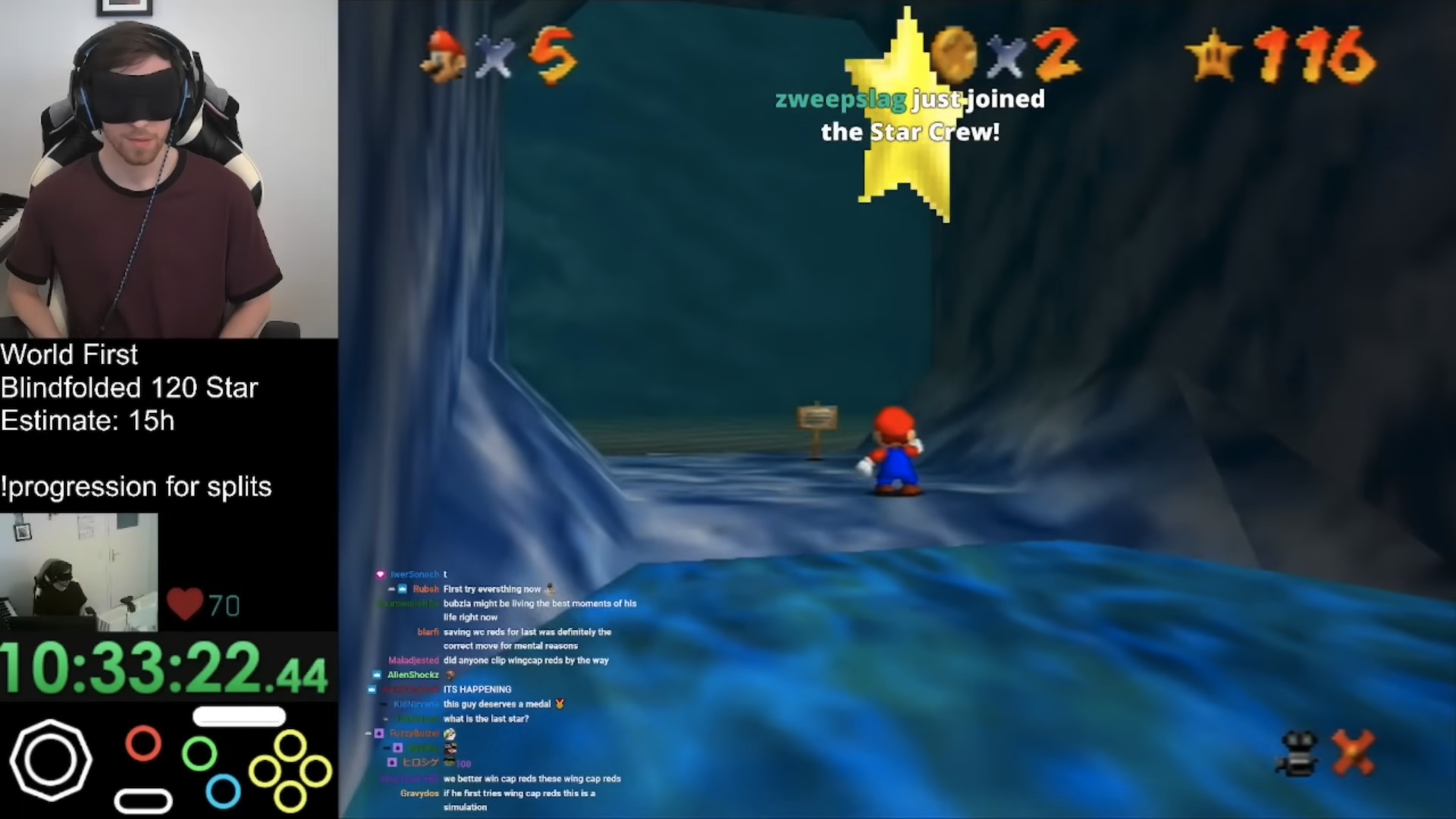}
		\caption{Bubzia}
	\end{subfigure}
	\caption{Captures of blindfolded speedruns by two gamers illustrating examples of ludic metagame in practice. (a) \textit{GiantGrantGames} performing a real-time strategy challenge in \textit{StarCraft II}~\cite{refGGGBlindfold}, and (b) \textit{Bubzia} executing a blindfolded \textit{Super Mario 64} speedrun~\cite{Bubzia11hour}. In both cases, the player relies solely on audio cues and memorised in-game timings and structures -- demonstrating an extreme form of ludic metagame where mechanical mastery intersects with creative self-imposed constraints.}
	\Description{Two sub-figures: (a) GiantGrantGames sits at a microphone wearing a blindfold and a headset, with a StarCraft UI on background; (b) Bubzia wears a blindfold while playing Super Mario 64, with the game screen showing Mario near a Power Star at the 116-star mark during a 120-star blindfolded run.}
	\label{fig:MetagameFig1}
\end{figure*}

For instance, a content creator named \textit{GiantGrantGames} has undertaken a variety of self-imposed challenges in the \textit{StarCraft} series -- deathless~\cite{refGGGDeathless}, no-build campaigns~\cite{refGGGnoBuild}, and even blindfolded playthroughs~\cite{refGGGBlindfold}. 
Similarly, \textit{Bubzia} performed a blindfolded speedrun of \textit{Super Mario 64} over eleven hours~\cite{Bubzia11hour}. 
As illustrated in Figure~\ref{fig:MetagameFig1}, such constraints -- while seemingly impossible -- push players to creatively exploit or repurpose game systems to succeed.

A widely recognised form of ludic metagame is the speedrun~\cite{refMHCodeIsLaw} -- the art of completing a game as quickly as possible~\cite{CambridgeDefSpeedrun, SoeedRundotComDef}.
Unlike mechanism-specific constraints (\textit{e.g.}, pacifist~\cite{MagooFo4Pacifist}, lowest score~\cite{CeaveMarioLowest}), speedrun centres on time as the primary metric of success, making it broadly applicable across genres.
These communities have evolved from niche online forums into globally recognised cultural phenomena.
Events like \textit{Games Done Quick (GDQ)} attract immense viewership; in 2020, \textit{GDQ}'s charity marathon peaked at 237,523 concurrent viewers -- 75.9\% of the \textit{Overwatch} League Grand Finals audience \cite{refMHCodeIsLaw}.
Such comparisons underscore that speedrunning is not merely a fringe activity but a thriving, competitive, and highly visible space.
Also, speedrunning is a semi-professionalised activity, as Hemmingsen notes~\cite{refMHCodeIsLaw}.

Speedrun culture is sustained by shared knowledge, route-finding, and peer-based verification~\cite{refBorn2Run, SpeedRundotComModeration}.
Strategy development and validation often take place across forums, Discord servers, and video platforms.
As such, speedrun is not just a solitary test of dexterity or memorisation -- it is a collaborative enterprise grounded in experimentation, documentation, and trust.

\subsection{Vulnerability in Ludic Metagaming Culture}
Increased visibility introduces risk. 
As streaming revenue, sponsorships, and social capital become increasingly accessible, some individuals have resorted to deceptive shortcuts -- including staged setups, plagiarised footage, or spliced runs. 
Forged or manipulated submissions exploit this trust-based ecosystem~\cite{refEZScapeBaboon, refAsmongoldFeud, AbyOsu, refDwangoDiablo, refAbyssoftMekarazium}. 
Even some of the respected figures in communities have been implicated \cite{refGhostSchmooey, refDreamIGN}. 
Further details of fraudulent cases are available in the appendix. 
As Snyder observes, “The real challenge of speedrun lies not in mastering reflex-heavy tricks or memorising multi-hour routes but in overcoming the psychological toll of resetting”~\cite{refMHCodeIsLaw}.

Criminological theories help explain the mechanism of ludic metagaming manipulation. 
The \textit{Routine Activity Approach} outlines that ``Crime is likely when three elements converge: (a) a motivated offender, (b) a suitable target, and (c) the absence of a capable guardian''~\cite{refMaybeRAA}. 
All three elements are clearly present in the context of manipulated speedruns: offenders are often skilled and motivated, digital footage is inherently malleable; and decentralised moderation lacks both the assistive tools and consistency for enforcement. 
\textit{Deterrence Theory} further argues that misconduct is deterred only when all three elements are present: (a) certainty, (b) celerity, and (c) severity of punishment \cite{refDeterentGibbs}. 
When any of these is absent, the likelihood of transgression increases. In online ludic metagaming communities, not just one -- but all three -- are lacking: 
detection is uncertain, responses are slow, and consequences are minimal, especially since legal action is rarely pursued.
This combination creates an environment unusually conducive, almost ideal, to misconduct.
Lastly, \textit{Rational Choice Theory}~\cite{refAkers, refCrimEco} posits that when the perceived benefits outweigh the risks of being caught, even actors with no prior misconduct may be tempted to commit crime. 
A community ban is often the harshest available penalty.
If the offender has not violated a specific law -- such as copyright infringement through stolen content -- there is often no recourse beyond public disapproval. 
Furthermore, even when laws are broken, enforcement is rare if the offence is a misdemeanour. 
In the case of the Badabun scandal, who plagiarised a speedrun by splicing together others' footage, the video remains publicly accessible years later, with no legal action taken despite clear violations.

\subsection{\textsc{Tracer}: A Forensic Detection Framework} 
To this end, we propose \textsc{Tracer}, \textit{Tamper Recognition via Analysis of Continuity and Events in game Runs}, a modular forensic framework for evaluation in speedrun submissions. This work makes the following contributions:
\begin{itemize}
   \item \textbf{First}, this paper presents a structured academic approach to systematically investigate fraudulent activities within ludic metagaming communities, moving beyond anecdotal documentation.
    \item \textbf{Second}, we conduct qualitative research, including audiovisual forensics, physical laws, human and hardware limitations, and cyberspace anomaly detection to ground our framework in both theoretical rigour and practical relevance.
    \item \textbf{Third}, we offer a generalisable set of evaluative heuristics for game replay analysis. While not automated, these structured guidelines can lower the cost of manual review and lay the foundation for future tooling.
\end{itemize}

\section{Related Works}
The application of forensic analysis to speedrun footage draws from multiple research domains--particularly media studies (as current gameplay records are in the form of HCI log), digital forensics, and online game security, from which we can adapt methods of automated anomaly detection that rely on normalised user input. 
The following lines of research provide conceptual and methodological foundations for structured manipulation detection.

\paragraph{Ludic Metagame and Cheat in Games}
Hemmingsen's work provides a theoretical foundation for understanding how player behaviour transcends formal game mechanics \cite{refMHLudicMetagame, refMHCodeIsLaw}. 
Brewer's grounded theory analysis of the speedrun community adds empirical weight, highlighting how notions of cheating are internally negotiated \cite{refBorn2Run}. 
Haapaniemi's thesis surveys detection and prevention strategies across a variety of genres \cite{refBachelorOulu}. 
Lastly, Scully-Blaker situates speedrun as a curatorial and performative act, where runners re-contextualise play through community-vetted norms, rule-sets, and recorded artefacts \cite{refScully}.

\paragraph{Media Forensics}
Fontani \textit{et al}.\ highlight cross-modal approaches for identifying digital splicing through inconsistencies introduced by recompression, frame insertion, or quantisation artefacts across image, video, and audio signals \cite{refFontaniSplice}.
In the audio domain, Jadhav and Vyas demonstrate the effectiveness of convolutional neural networks applied directly to spectrograms for detecting inserted segments \cite{refJadhavAudioCNN}. 
For visual media, Tiwari and Dixit propose a Mask-RCNN-based method that localises spatial anomalies consistent with copy-paste manipulations \cite{refTiwariRCNN}.
    
\paragraph{Log-Based Game Bot Detection} 
In order to classify cheating users from normal users, feature selection and machine learning or deep learning are applied at the server-side in general. 
Server-side cheating detection based on user's play log are long been accepted by industry practitioners such as \textit{NGUARD}, a generalised game bot detection framework~\cite{refNguard}, chatting log based detection~\cite{refHKBotAreumTalk}, specific play style based detection \cite{refHKBotAreumParty}.
Their work, using a log-based automated detection tool, can be adopted to \textsc{Tracer} to automate the process once the game player's input data is normalised. 

Together, these domains validate the use of structured, forensic methodologies to assess the authenticity of gameplay footage, making \textsc{Tracer} a synthesis of media and behavioural security practices.

\section{Threat Model}\label{sec:threat-model}
\subsection{Taxonomy of Interaction Spaces}

\begin{figure}[t]
	\centering
	\includegraphics[width=\linewidth]{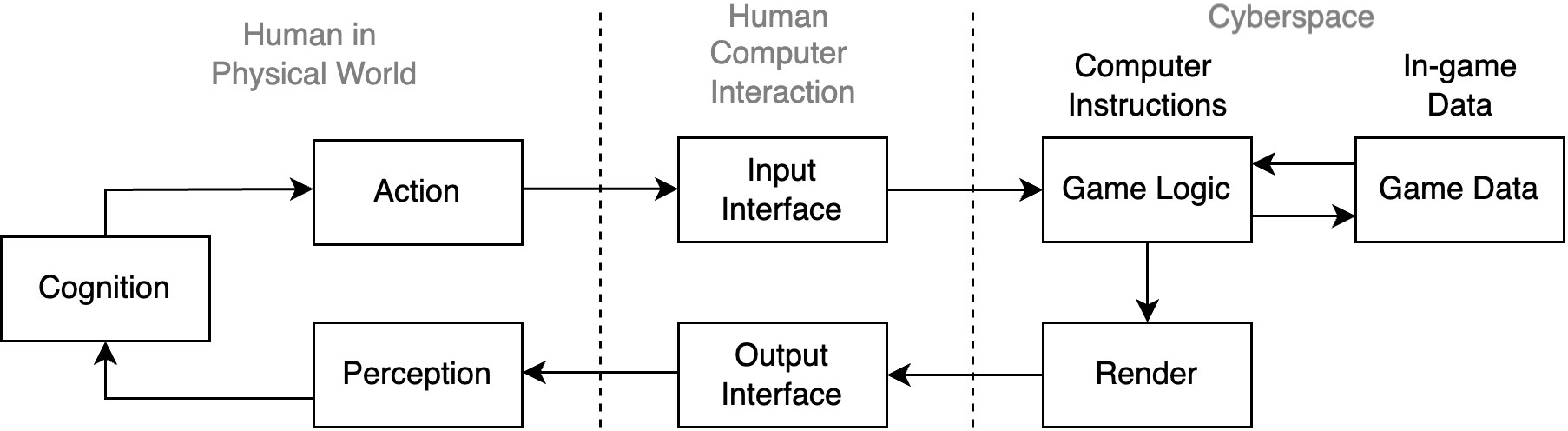}
	\caption{A conceptual model of the human-computer interaction (HCI) loop in gameplay. The system is divided into three distinct regions representing potential attack surfaces: the physical world (\textit{e.g.}, laws of physics, human motor/cognitive limitations), the HCI boundary (input/output interfaces and media), and cyberspace (game logic, data, and engine state). Each region corresponds to unique attack vectors for manipulation in gaming scenarios.}
	\Description{A flow diagram of the human-computer interaction loop in gaming, divided into three sections: Human, Human-Computer Interaction, and Computer. On the human side, cognition leads to action, which is passed to the Input Interface. In the computer section, the Input Interface sends data to Game Logic, which reads from and writes to Game Data. Game Logic also sends instructions to the Render module. The rendered output goes through the Output Interface back to the human via perception, completing the loop.}
	\label{fig:hci-loop}
\end{figure}

To understand how manipulation occurs in ludic metagaming communities, we first define the key domains where each important component is situated within the gameplay feedback loop.
Figure~\ref{fig:hci-loop} illustrates this loop, divided into three conceptual domains: 
(1) the Physical World, (2) the HCI Interface, and (3) Cyberspace.

\subsubsection{\textbf{Physical World}}
This domain encompasses the player's body, their immediate surroundings, and the physical hardware used to interface with the game.
It represents the foundational layer of gameplay interaction -- where biological capability, mechanical responsiveness, and physical law collectively define the boundary of legitimate input that should not be violated in authentic gameplay.

\paragraph{(a) Physiological Limits.}
This subdomain concerns the biological constraints of players, including reaction time, motor coordination, and cognitive load.
For instance, e-sports athletes can exhibit visual reaction times in the range of 150–300 milliseconds \cite{refEsportReactionTime} and in rare cases of sprinting, reportedly dipping below the 100ms threshold \cite{refSub100ReactionPossible, kiltyTweet2021}.
This sets a realistic benchmark for what is physically achievable. Repeated sub-millisecond actions, or executing more than ten discrete finger inputs simultaneously, exceed typical human capability and therefore raise questions of authenticity.
Interpretations of player performance must account for such physiological limits.
    
\paragraph{(b) Hardware Constraints.}
The physical characteristics and performance limitations of gaming equipment also shape what is possible during gameplay.
For instance most computer keyboards have limited key rollover, making more than a certain amount of simultaneous inputs infeasible.
Likewise, byproducts such as mechanical latency, controller delay, or analog noise (\textit{e.g.}, 50/60 Hz power-line hum) are notable artefacts.
These hardware signals, though captured digitally, originate in the physical domain and often serve as contextual markers of authenticity.
    
\paragraph{(c) Physical Law Boundaries.} 
This subdomain includes constraints grounded in fundamental physics -- covering motion, timing, and causality. 
The plausibility of any performance is ultimately bounded by these principles. 
Examples include the impossibility of simultaneous body movement and gameplay control, or the absence of physical interaction during input-dependent sequences. 
Interpretations at this layer require careful attention to the alignment of physical and in-game behaviours.

\subsubsection{\textbf{HCI Interface}}
The Human-Computer Interaction (HCI) layer forms the boundary between the physical world and the digital game environment. It includes all input and output channels through which gameplay is mediated: keyboards, mice, controllers, capture cards, monitors, and screen overlays.

This layer is especially relevant for verification, as the majority of play record -- including handcam footage, controller telemetry, and visual overlays -- originates here though the target information is in a different domain. 
Discrepancies between physical behaviour and digital response often reveal manipulation. 
However, this layer can be vulnerable to desynchronisation \cite{refAbyssoftMekarazium}, selective capture \cite{refAbyDiablo}, or post-processing tampering \cite{refAbysSchmooey, refKJBaboon, refSeastarYap}, making it a critical zone for forensic scrutiny as shown in each case.

\subsubsection{\textbf{Cyberspace}}
Cyberspace refers to the in-game environment itself -- the executable logic, runtime state, and in-game data. 
Unlike the physical world, this virtual space is malleable. 
Artificial data and execution logic can be injected, modified, or overwritten without any physical trace. 
Thus, recognising tampering via evidence can be difficult and sometimes impossible.

Threats here include in-game mechanic -- such as pseudo random number generation (PRNG) oracle -- manipulation, memory editing, run splicing, speed modifications, and injected scripts. 
Because this layer operates invisibly to the observer, it relies heavily on indirect cues (\textit{e.g.}, statistically improbable outcomes or illogical state transitions) for detection. 
Of the three domains, cyberspace offers the greatest flexibility to adversaries but also the richest context for forensic inference when anomalies arise.

\paragraph{(a) In-game Data}
In-game data refers to all information structures that represent the game's internal state, including player position, inventory, timers, flags, event triggers, and variable values. 
This data persists in memory and forms the substrate upon which gameplay unfolds. 
While often invisible to the player, it can be read, altered, or injected into memory manipulation, save file editing, or script-based tampering. 
In speedrun contexts, manipulated in-game data may manifest as suspicious item availability, inconsistent state flags, or abrupt transitions not explainable by normal gameplay mechanics. 
Although game data is inherently non-visual, its effects become visible through gameplay behaviour and outcomes.

\paragraph{(b) In-game Logic}
In-game logic refers to the rule-based systems and computer instructions that govern how game states evolve in response to inputs and internal conditions. 
This includes event handling, physics simulations, game mechanics (\textit{e.g.}, damage calculation, collision detection), and progression logic. 
Unlike raw data, logic defines the relationships between data points and determines how they change over time. Manipulating game logic -- for example, through code injection, runtime patching, or abuse of unintended interactions -- can produce outcomes that appear valid on the surface but violate the intended flow of gameplay. 
Anomalies such as inconsistent behaviour under identical inputs, unexpected bypassing of scripted events, or deterministic abuse of pseudo-random mechanics often signal tampering at the logic level.

\paragraph{(c) Rendering}
The render subdomain encompasses the processes responsible for generating the audiovisual output seen by the player -- including graphics rendering, audio synthesis, camera positioning, UI overlays, and frame composition. 
While render systems do not directly influence gameplay mechanics in many cases, they define how game state is presented to the observer and are thus critical in forensic evaluation. 
Manipulations at this level include frame injection, selective overlaying, resolution or framerate spoofing, and post-production edits. 
Because the rendered output is often the only observable layer in video submissions, discrepancies between what is rendered and what should logically occur can be strong indicators of tampering. 
Forensic integrity often hinges on consistency between render output and lower-layer state transitions.

\subsection{Threats on Attack Surface}
\subsubsection{\textbf{Physical World}}
Threats within the physical domain target what is plausible given human biology, hardware limitations, and physical laws. While such manipulations are harder to automate, they often exploit observers' assumptions or rely on subtle staging. We organise threats here into three subcategories:

\paragraph{(a) Human Limitation Masquerade.}
Some manipulations aim to fabricate performance beyond plausible human ability. 
For example, sequences requiring sub-millisecond reaction times or simultaneous input across more keys than physically reachable suggest the use of input automation or pre-recorded macros~\cite{AbyOsu}. 
Even when handcam footage is present, subtle signs -- such as unnatural consistency or mechanical smoothness -- may indicate human implausibility. 
In \textit{Queen Pwnzalot}'s case, a blindfolded challenge was undermined by a suspicious seating position and the presence of a secondary, off-screen monitor -- suggesting retained visual access. 
Such staging exploits assumptions about sensory deprivation and environment control.

\paragraph{(b) Hardware Deception.}
Many frauds exploit overlooked properties of physical input devices. Mechanical noise from peripherals (\textit{e.g.}, keypress clicks, controller vibrations, microphone hum) often leaves artefacts in recordings~\cite{refKJSeastar}; the absence or mismatch of such cues can signal tampering or post-production manipulation.

\paragraph{(c) Physical Law Violations.}
Some behaviours contradict physical causality altogether. 
For instance, if a player is seen scratching their head while the in-game character is actively controlled, as shown in \textit{Mekarazium}'s case \cite{refKJMekarazium}, the footage likely involves pre-recording or desynchronised input feeds. 
In \textit{Schmoomey}'s case, some of the uploaded videos were sped up in post-production, resulting in discrepancies in offender's bodily movement \cite{refAbysSchmooey, refGhostSchmooey}.

\paragraph{Input Evaluation as Forensic Practice.}
One of the key detection practices in the community is input evaluation: the process of analysing whether a player's physical input plausibly aligns with the in-game response.
Evidence sources include handcam footage, on-screen input overlays, and controller telemetry.
However, verification is often hindered by poor lighting, low resolution, compression artefacts, or obscured finger movements.
Investigators may scrutinise not only the timing of button presses but also the feasibility of the hand techniques involved -- particularly in rhythm games like \textit{Osu!} or \textit{Geometry Dash}, where precise timing and motion are critical.
Discrepancies between the expected and observed motor actions often serve as the definitive indicators for fraud.

\subsubsection{\textbf{HCI Interface}}
Some game titles, such as \textit{Doom} or \textit{TrackMania}, provide native input-logging tools, allowing replays to be shared in lightweight formats. However, in most cases, players log their gameplay into full audiovisual recordings.
These outputs effectively serve as ``HCI logs'' -- records of human-machine interaction -- and are therefore prime targets for manipulation.
Compared to other vectors, HCI manipulation does not require specialised hardware, as physical attacks do, nor does it demand deep technical expertise like memory-level exploits in cyberspace. 
This makes it a broadly accessible and attractive vector for malicious actors.

Common methods include video editing, input simulation via emulated devices, and log tampering.
For example, keyboards with macro functionality can simulate highly complex and precise input sequences.
More advanced attackers may use microcontroller-based injection tools to imitate controller signals -- a method feasible with basic embedded systems knowledge. 
Because of these threats, many speedrun and challenge communities now require handcam or facecam footage to accompany submissions -- aiming to restore trust through redundant verification layers.

\paragraph{Screen Evaluation.}
Moderators scrutinise gameplay footage for visual artefacts that may suggest tampering. 
Indicators include abrupt lighting changes, UI mismatches, or scene transitions that reveal splice points -- often betrayed by transparency layers, HUD misalignment, or unexpected cropping. 
Such anomalies are commonly associated with time-wise edits or compositing of separate play segments.

\paragraph{Audio Evaluation.}
Audio channels can reveal manipulation that visuals fail to show.
Reviewers analyse waveform continuity, background ambience, and alignment between game events and sound cues (\textit{e.g.}, keystrokes or button presses).
In several known cases, fraudulent submissions involved mixing separately recorded gameplay and commentary tracks to mask pre-recorded inputs.
Tell-tale signs include phase inconsistencies, abrupt ambient shifts, or mismatched reverb profiles between scenes.

\subsubsection{\textbf{Cyberspace}}
This domain refers to the internal logic, memory states, game save, and system behaviour encoded within the game software itself.
Unlike physical and HCI layers, where manipulation targets observable interaction, cyberspace attacks modify or exploit the digital substrate that governs gameplay mechanics and data.

\paragraph{System-Level Exploitation.}
This includes memory editing, script injection, or replay splicing at the file level. 
Attackers may alter game data directly -- bypassing the need for observable input/output manipulation. 
These attacks are harder to detect without deep engine knowledge or integrity verification tools.

\paragraph{Game Engine Specificity.}
Each game has built-in behavioural patterns and execution rules. Experienced moderators leverage these to detect irregularities. 
For instance, in \textit{Super Mario 64}, creatures' eye-blink pattern is synchronised with exact timing, making unsuspecting offenders' splicing detectable~\cite{Mario64BlinkTwoSecs}.
In \textit{Jump King}, in-game timers must align with specific tick intervals~\cite{refKJSeastar}. 
Violations of such internal consistency can be detected by simple automated scan and output timing anomalies.

\paragraph{Contextual Gameplay Analysis.}
Moderators compare observed gameplay with expected in-game statistics. 
For example, kill counts, item pickups, and quest states that don't align with the game's known logic may indicate manipulation -- whether through direct save editing or record splicing. 
Some forms of manipulation, especially in carefully forged video logs, may leave no visual splice traces. In such cases, contextual anomalies become critical evidence.

\paragraph{Probabilistic Inconsistencies.}
Games with randomised mechanics can be reverse-analysed for statistical plausibility. 
For instance, \textit{Minecraft} world generation and loot drop mechanics are determined by seeded PRNGs~\cite{minecraftSeedWiki}. 
\textit{Dream}'s infamous speedrun case was debunked by demonstrating that his success rates on item barters were statistically implausible compared to known internal rate -- even after accounting for variance~\cite{refKJDream1}.
Similarly, in \textit{Diablo}, one debunked run was flagged due to impossible map and quest combinations generated from reverse-engineered seeds~\cite{refDwangoDiablo}.

\section{\textsc{Tracer} Framework}
\subsection{Forensic Process Pipeline}
\begin{enumerate}
	\item \textbf{Evidence Acquisition.} Collect raw evidence such as video recordings, log files, and configuration files. For certain titles like \textit{Minecraft}, moderators do acquire the player's world data to verify procedural generation consistency in specific cases.
	\item \textbf{Artefact Normalisation.} Convert heterogeneous artefacts into structured representations. For example, extract frame-level input timestamps, and quantify game-specific metrics such as barter count or movement paths.
	\item \textbf{Event Attribution.} Map observed events into their corresponding forensic domains (physical, HCI, cyberspace) to prepare for dimensional analysis.
	\item \textbf{Analytical Review.} Use domain-specific heuristics to identify inconsistencies or manipulative artefacts. Community-driven evaluation can operate in parallel to triangulate uncertain cases.
	\item \textbf{Verdict Synthesis.} Provide a formalised verdict. In community settings, this step feeds into moderation workflows; in governed contexts, it may support committee decisions or rule enforcement.
\end{enumerate}

\subsection{Analysis module}
Tracer, as a set of countermeasures of the threats shown in \S~\ref{sec:threat-model}, it organises detection strategies into three analytical dimensions: \textbf{Physics Coherence}, \textbf{Media Continuity}, and \textbf{Cyberspace Consistency}. 
Each module offers domain-specific heuristics that can be applied independently or jointly to assess the authenticity of gameplay footage and associated metadata.

\subsubsection{Physics Coherence Module}
This module evaluates whether observed player inputs and execution patterns are plausible within human and hardware constraints. 
It focuses on physical and bio-mechanical feasibility, drawing from known reaction times, key rollover limitations, and other empirical performance benchmarks. Key indicators include:

\begin{itemize}
	\item Implausible input sequences, such as timing below human reaction thresholds or exceeding hardware limit.
	\item Mismatches between handcam footage or input overlays, and in-game outcomes that suggest pre-recording or automation, or separately recorded handcam and gameplay merged into one video.
	\item Unjustified deviations from established strategy or execution style that lack in-run justification, possibly indicating splicing or tool-assisted segments.
\end{itemize}

This module is particularly effective against submissions that reconstruct human input through staged footage or simulation. 
It encourages verification against known physical markers, including muscle fatigue patterns and keypress artefacts captured in ambient audio or video.

\subsubsection{Media Continuity Module}
This module targets audiovisual inconsistencies resulting from footage manipulation, including splicing, compositing, or timing mismatches. It applies frame-level inspection and forensic media analysis to detect irregular transitions, altered overlays, or temporal discontinuities. Core heuristics include:
\begin{itemize}
    \item Frame-by-frame examination for abrupt scene transitions, jittering HUD elements, or tonal inconsistencies.
    \item Desynchronisation between audio and video streams -- for example, sounds that do not align with visual input.
    \item Shifts in ambient audio that suggest multiple recording sources or post-production mixing.
\end{itemize}
These cues are adapted from splice detection literature and digital signal analysis. Gameplay footage is especially susceptible to compression artefacts, making redundancy in audio-visual alignment a useful forensic feature.

\subsubsection{Cyberspace Consistency Module}
This module evaluates logical consistency within the game's internal environment. It considers whether game states, statistics, and progression align with known mechanics, scripted events, and platform-specific behaviour. Red flags include:
\begin{itemize}
    \item Statistical or logical anomalies across inventory, health, or environmental state between checkpoints or frames.
    \item Anomalies in pseudorandom behaviour (\textit{e.g.}, drop rates, map generation) that significantly diverge from the game's internal probabilities.
    \item Absence of expected triggers, dialogue, or transition animations associated with normal progression.
\end{itemize}

For titles with deterministic PRNG or strict event sequences, the module also supports probabilistic analysis. Cases such as \textit{Dream}'s \textit{Minecraft} one \cite{refDreamPolygonSus} and \textit{Diablo} one \cite{refDwangoDiablo} demonstrate the efficacy of reverse-analyzing PRNG logs to expose manipulated outcomes.

\section{Discussion}
\subsection{Limitations}
As this work is submitted to a \textit{Work-in-Progress} track, the proposed framework remains under active development.
In addition, the current implementation is a manual framework, relying upon human judgment and forensic inspection.
Although the process is structured and repeatable, it remains inherently labour-intensive, which still limits its scalability in high-volume review contexts.

\subsection{Future Work}
We plan to mature the framework by incorporating more formalised criteria and validated use cases. 
As more annotated data becomes available, we aim to strengthen both the structure and practical applicability of \textsc{Tracer}.
Additionally, we intend to formalise case studies, both previously debunked cases and synthetic (hypothetical but plausible scenarios), to support training and calibration.
Providing such examples may help moderators and community members improve both preparedness and evaluation accuracy.

Building on \textsc{Tracer} as a foundational evaluation process, multiple directions for automation can be pursued.
First, audiovisual forensic automation may be developed or adapted from existing tools to detect anomalies in gameplay footage. 
Second, gameplay logs can be harvested once a suitable logging mechanism (\textit{e.g.}, input or memory trace collector) is developed and adopted by ludic metagaming communities.
The logs could then serve as input for automated anomaly detection models, drawing on methods from bot detection.
While data sparsity in this context may hinder full automation, assistive AI or machine learning-based tools would still substantially improve evaluation efficiency.

\appendix
\section*{Appendix (Formerly Supplementary Material)}
\addcontentsline{toc}{section}{Appendix}
\section*{Additional Examples and Details}

This appendix brings together the examples and case details originally published as supplementary material for our CHI PLAY 2025 Work-in-Progress paper~\cite{yoo2025tracer}.

\section{Known Incidents of Forged Ludic Metagame Runs}
While many participants in ludic metagaming communities act in good faith, there have been numerous cases where individuals submitted forged or manipulated runs. 
However, there have also been known instances of accusations widely regarded as false \cite{refAbysForensic}, prompting caution in passing judgment. 
Since absolute human intent is unknowable, this material represents on cases where (a) the player admitted manipulation, (b) the accused effectively vanished from the community without refuting credible evidence, or (c) a verdict was issued by a respected authority (\textit{e.g.}, moderation committees of \textit{Games Done Quick} or platform-specific bodies). 
The subsequent incidents underscore the need for more systematic and transparent verification mechanisms. 
By examining the following five exemplary cases, we can better understand common manipulation techniques and how they were detected.

\subsection{\textit{Badabun}'s Plagiarized \textit{Super Mario} Run}
\begin{figure} [h]
	\centering
	\includegraphics[width=\linewidth]{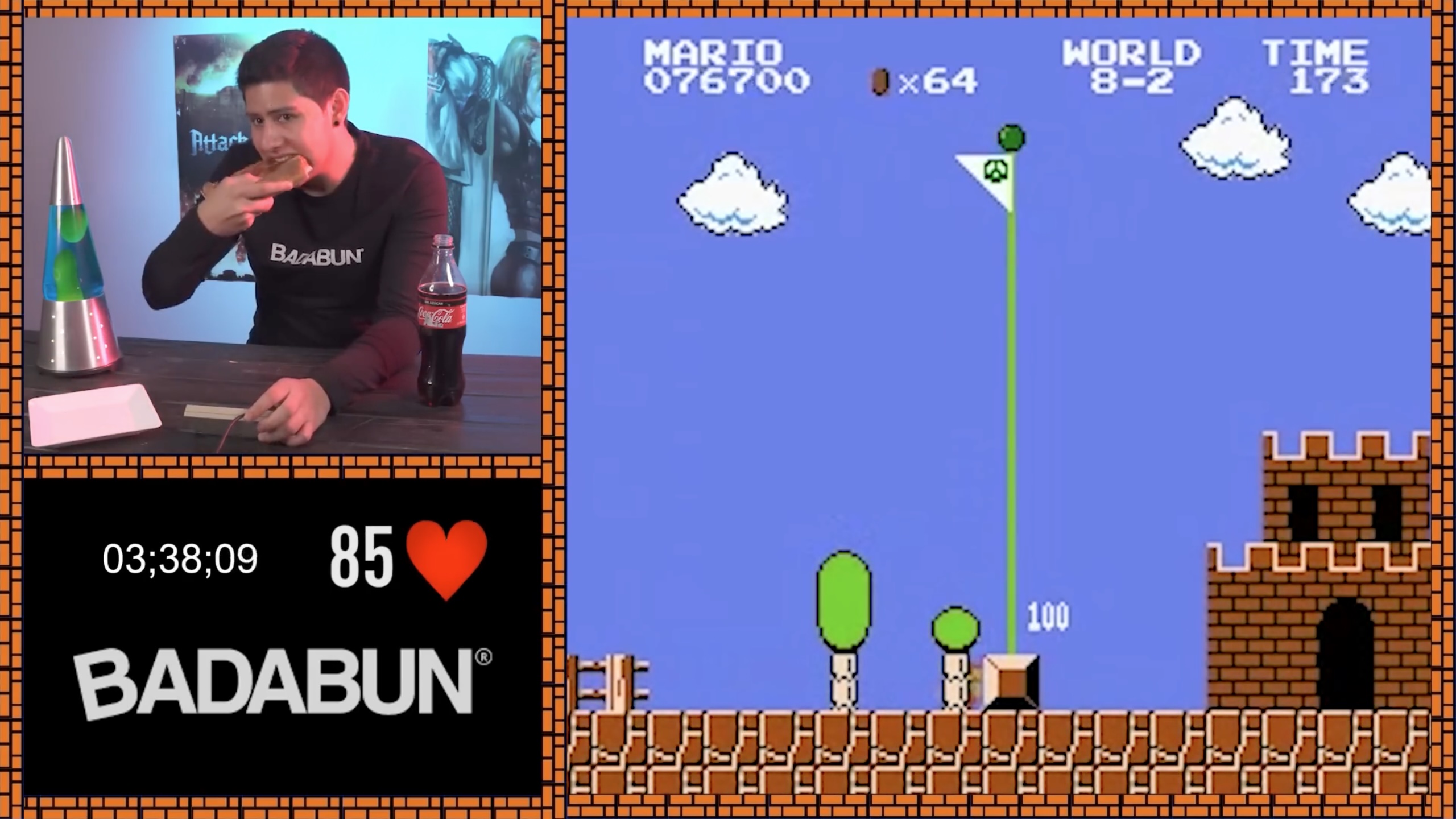}
	\caption{\textit{Gustavo}, the player shown in \textit{Badabun} channel's video, eating pizza during his alleged \textit{Super Mario Bros.} speedrun~\cite{refBaboon}. The casual setup contributed to community skepticism regarding the run's authenticity.}
	\label{fig:baboon}
	\Description{Split-screen layout. Left side shows a man eating pizza while holding a game controller, with a beverage bottle on the table. Below him are a timer, a heart rate reading of 85, and the logo "Badabun." Right side shows gameplay from Super Mario Bros, with Mario reaching the flagpole in World 8-2.}
\end{figure}
One of the most notable cases of speedrun fraud came from \textit{Badabun}, a Spanish-language YouTube channel. 
In 2017, they uploaded a video claiming to have achieved an exceptional time in \textit{Super Mario Bros.}~\cite{refBaboon}. 
However, the run was fabricated from the ground up -- composed of stolen segments from legitimate players, edited together as a continuous run~\cite{refKJBaboon}. 
Discrepancies in in-game audio and transitions revealed the run was not authentic. 
Blatant inconsistencies such as score tracking further exposed the inept editing and a fundamental misunderstanding of the game's mechanics and speedrun conventions. 
His casual demeanour -- eating pizza, drinking beverage and featuring a feign heart rate monitor oscillating between two numbers alienated the speedrun community and even became the subject of widespread ridicule and internet memes~\cite{refEZScapeBaboon, refbaboonParady}.
Noting that despite the evident fraud, the video remains online and unlabelled, continuing to generate viewership. 
They later uploaded an apologetic video, describing the fabricated speedrun video as a harmless prank and insisting that no one was supposed to be misled~\cite{refEZScapeBaboon}. 
As of 22 July 2025, the apology video is no longer publicly available; however, a translated mirror remains accessible via a third-party upload~\cite{refSozTran}.

\subsection{\textit{Schmooey} and the \textit{Guitar Hero} Scandal}
\begin{figure} [h]
	\centering
	\includegraphics[width=\linewidth]{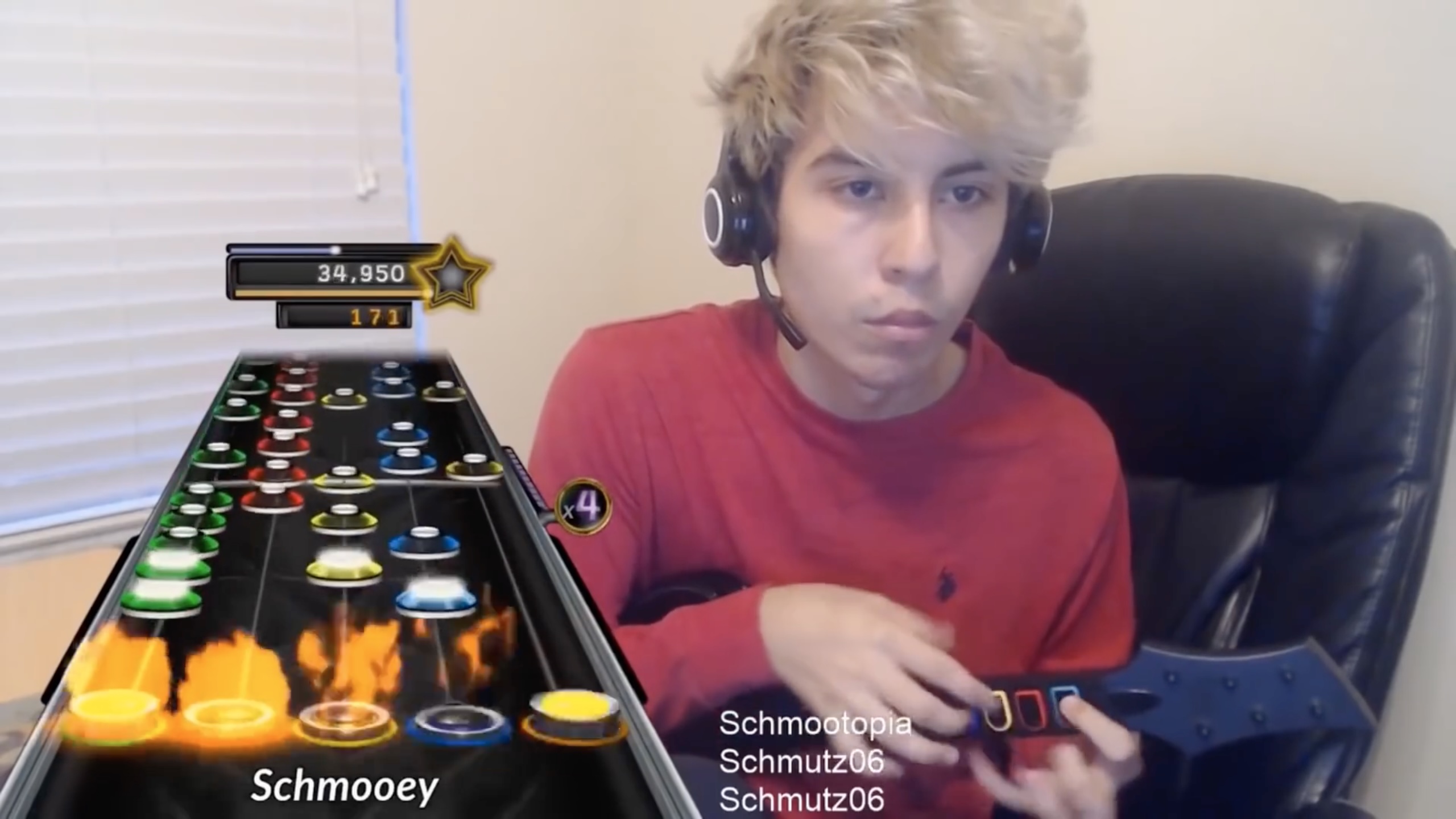}
	\caption{A player named \textit{Schmooey} performing a \textit{Guitar Hero} run~\cite{refGhostSchmooey}. Once a respected figure in the community, he was later revealed to have manipulated many of his high-profile submissions.}
	\Description{A man wearing a headset holds a plastic guitar-shaped controller, pressing its buttons while focusing intently on the screen. On the left side of the image, an overlay of Guitar Hero gameplay is shown, with densely packed coloured notes, a score display, a multiplier, and flames beneath the note track. The player name 'Schmooey' is visible.}
	\label{fig:schm}
\end{figure}
\textit{Guitar Hero} is a rhythm-based music game where players simulate playing guitar by pressing buttons in sync with on-screen prompts.
\textit{Schmooey} was a well-known and active member of the \textit{Guitar Hero} community. He frequently participated in community meet-ups, co-played with other players, and was widely regarded as highly skilled and experienced.
He gained further acclaim for his apparent ability to full-combo tracks (a run without a single missing note) that even top-tier players found nearly impossible, and regularly claimed bounties for “first-ever” clears.

Over time, however, his videos began to draw skepticism due to their near-perfect consistency, lack of mistakes, and suspicious visual artefacts. 
Investigations revealed that many of his submissions were heavily edited. 
In some cases, he used external tools such as Cheat Engine to time-stretch gameplay then record the gameplay in a slower pace. 
Later he sped the in-game footage back up to normal speed. 
He also recorded separate footage of himself miming gameplay and composited it onto falsified gameplay footage -- sometimes even showcasing physically implausible hand techniques that did not match the actual note timings. 
In other instances, he genuinely played the game with a facecam, but at reduced speed, later accelerating the entire video. 
This approach introduced sudden jolts and visual inconsistencies in the facecam footage \cite{refAbysSchmooey}.

After being confronted multiple times with the evidence, \textit{Schmooey} admitted to the fraud and withdrew from the community. 
As thoroughly documented by several community analysts and content creators -- including \textit{Karl Jobst, Abyssoft, and Ghostforce} -- who investigated and published detailed breakdowns of the case on YouTube \cite{refKJSchmooey, refAbysSchmooey, refGhostSchmooey}, the case stands out due to the respect \textit{Schmooey} had earned prior to the scandal. 
He had received nearly USD \$3,000 in community bounties, illustrating how well-intentioned incentive structures can be exploited. 

\subsection{\textit{EpicSeastar}'s Reconstructed \textit{Jump King} Speedrun}
\begin{figure} [h]
	\centering
	\includegraphics[width=\linewidth]{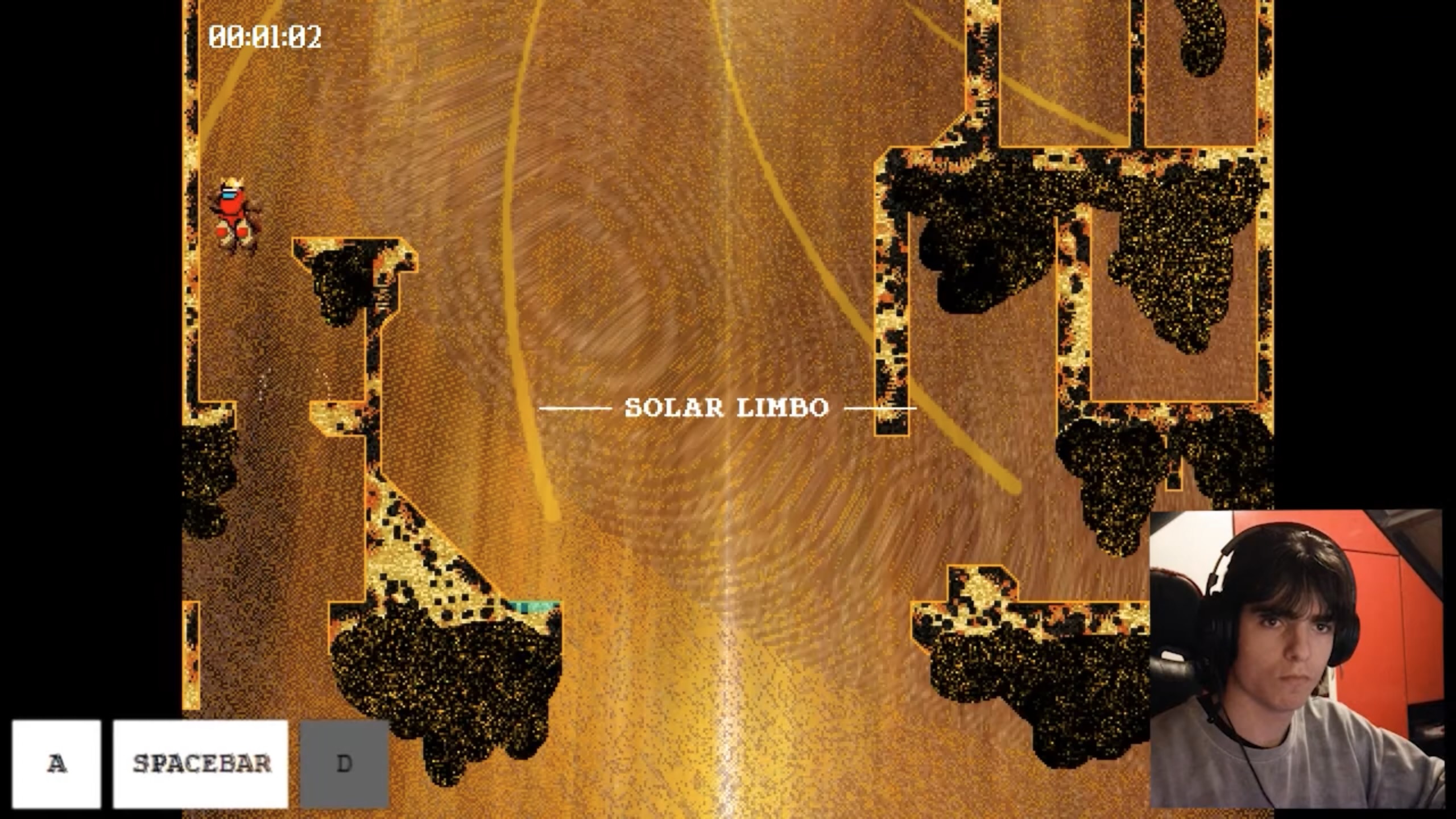}
	\caption{A fabricated \textit{Jump King} run uploaded by \textit{EpicSeastar}~\cite{refSeastarFake}. He heavily spliced footage to shorten the playtime and post-processed the video to recreate the in-game UI. He also recorded a staged reaction of joy and surprise to enhance the illusion.}
	\label{fig:seastar}
    \Description{A gameplay screen is shown, with a timer in the top-left corner and the name of the stage at the centre. In the bottom left corner, a keyboard input overlay displays which keys are being pressed. On the right side, a facecam shows EpicSeastar wearing a headset and focusing intently on the screen.}
\end{figure}
In 2022, a gamer named \textit{EpicSeastar} submitted a few \textit{Jump King} speedruns with completion times that reached world-record levels. 
\textit{Jump King} is a vertical platformer in which players ascend a tower using only charged jumps. 
While initially celebrated, frame-by-frame review revealed irregularities in scene transitions, background jitter, and jump momentum -- all strongly suggesting that the footage had been spliced. 
Further investigation uncovered that multiple audio channels were used, including a dedicated keyboard sound track intended to mimic real-time input, and that the on-screen keypress display had been manipulated to align with edited footage.

In addition to video editing, \textit{EpicSeastar} performed a staged reaction of joy and surprise, which was later scrutinised as part of the broader evidence of deception \cite{refKJSeastar}. 
Unlike some offenders who removed their content, one of the original videos remains publicly available and continues to collect views \cite{refSeastarFake}. 
He also published an explanation video shortly after the exposure, describing the submission as a prank intended to ``fool people first and then let them know it was fake'' \cite{refSeastarYap}, which he failed to do until the deception had been exposed.

The case underscores how elaborate editing and performative cues -- when left unchecked -- can produce convincing counterfeits.

\subsection{\textit{Queen Pwnzalot} and Staged Live Streams of Blindfolded \textit{Monster Hunter} Runs}
\begin{figure} [h]
	\centering
	\includegraphics[width=\linewidth]{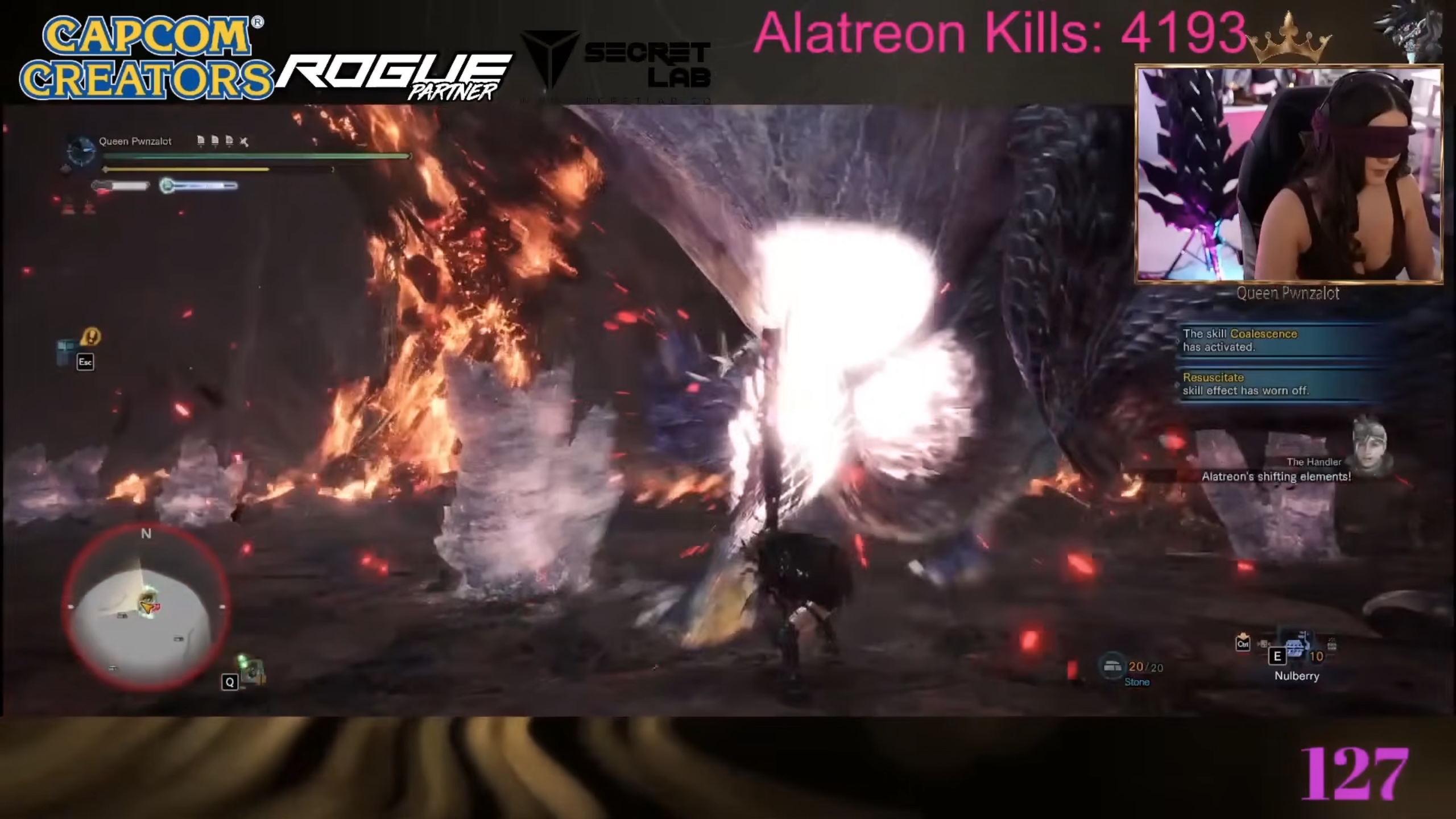}
	\caption{A streamer \textit{Queen Pwnzalot} playing \textit{Monster Hunter} while allegedly blindfolded~\cite{refKJQueen1}. Both runs were livestreamed, illustrating that real-time streaming platforms are not immune to attempted deception.}
	\label{fig:pwnz}
    \Description{Gameplay screen of Monster Hunter showing a battle against the monster Alatreon. In the top-right corner, a facecam shows a woman wearing a blindfold and a headset, seated in a gaming chair. Her username, Queen Pwnzalot, is displayed below. Text at the top reads "Alatreon Kills: 4193" and various HUD elements such as health bar, minimap, and item shortcuts are visible.}
\end{figure}
\textit{Queen Pwnzalot}, a Twitch live streamer, uploaded two videos of her live stream sessions claiming to complete \textit{Monster Hunter} gameplay while blindfolded.
\textit{Monster Hunter} is a third-person action game series, in which players hunt huge monsters. The gameplay emphasises precise timing, spatial awareness, and visual cues, making blindfolded play especially difficult.
Her claims drew intense scrutiny from the community, ultimately exposing multiple layers of deception. In the first video, she wore a visibly translucent blindfold, clearly compromising the legitimacy of the run as shown in~\cite{refKJQueen1}.
Based on audio cues, an analysis raised questions as to whether she had access to information beyond what was audible~\cite{refAbyssoftQueen1}.
The first video sparked significant controversy, prompting debate and conflict between content creators and their audiences, as documented in \cite{refAsmongoldFeud}.
In the second run, although she used a legitimate blindfold and appeared to cover her visible monitors with thick papers, investigators pieced together that a third, off-camera screen mirroring gameplay was used to mirror gameplay.
\cite{refKJQueen2} Before fully donning the blindfold, she responded to live chat despite allegedly lacking visual access. During gameplay, she notably avoided using auto-aim features, which are commonly relied upon in genuine blindfolded runs. Instead, she manually aimed while repeatedly looking upward to track the monster's head position -- offering evidence of visual feedback.
Further analysis, including forensic breakdowns by content creators, revealed how the entire setup had been staged to appear legitimate~\cite{refAsmongoldForensic, refAbyssoftQueen2}. As of 22 July 2025, \textit{Queen Pwnzalot}'s Twitch channel~\cite{refQwnTwitch} and her YouTube channel~\cite{refQwnYouTube} contains little to no publicly accessible contents, suggesting she has withdrawn from the gaming community.

While no post-editing was involved, the deception unfolded live -- a calculated performance exploiting viewer trust. This case exemplifies that even real-time content can be manipulated.

\subsection{\textit{Dream}'s Altered PRNG in \textit{Minecraft} Speedrun}
\begin{figure} [h]
	\centering
	\includegraphics[width=\linewidth]{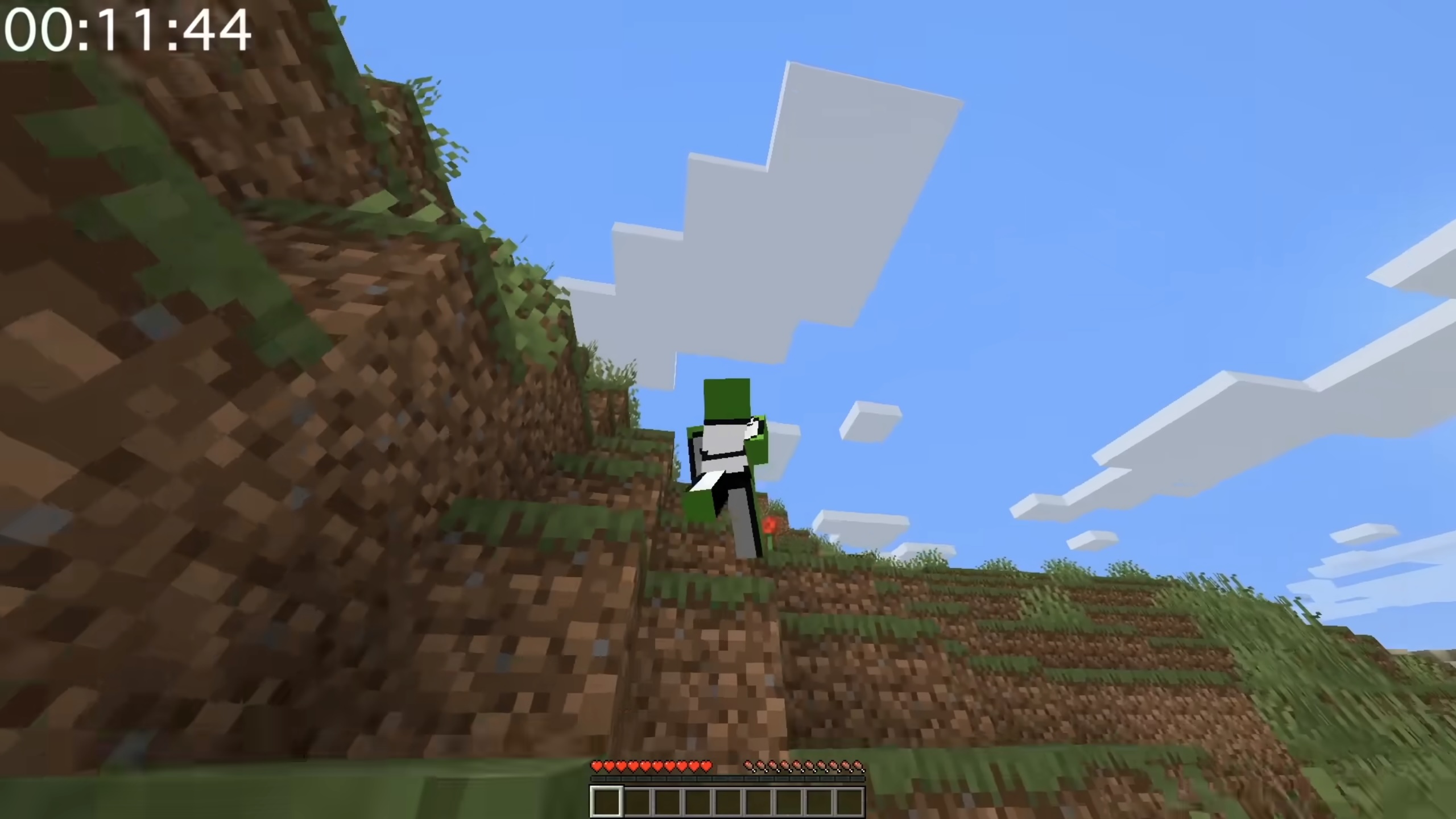}
	\caption{\textit{Dream}, a content creator and speedrunner, playing \textit{Minecraft} in version 1.14 \cite{refDreamSpdrun}. He later submitted a version 1.16 world record run that sparked allegations of manipulation -- which he eventually confessed to.}
	\label{fig:dream}
    \Description{A Minecraft character wearing Dream's in-game skin, which is green and white, is looking directly at the in-game camera while standing on a grassy hillside. A timer in the top-left corner reads 00:11:44. The in-game UI is visible at the bottom of the screen.}
\end{figure}
In 2020, a YouTube content creator known as \textit{Dream} submitted a \textit{Minecraft} speedrun that featured statistically improbable item drop rates -- notably for \textit{blaze rods} and \textit{ender pearls}, both of which are essential items to game completion.
Initial review placed the run within acceptable margins, but growing suspicion prompted a formal investigation by the \textit{Minecraft} speedrun moderation team~\cite{refDreamPolygonSus}. 
After extensive statistical analysis, they concluded the odds of the observed drops were highly unlikely and determined that the game's pseudo random number generation (PRNG) behaviour had been modified.
\textit{Dream} denied wrongdoing at first but later admitted to having run with altered game internals \cite{refKJDream1, refKJDream2, refDreamScreenrant, refDreamIGN, refDreamPCgamer, refDreamGameRant, refDreamInquirerDotNet}. 
The case sparked widespread debate, not just about integrity but about verification procedures and statistical evidence as given at \cite{refDreamModerator}. 
Notably, the incident was covered by both community analysts and media, such as \textit{The Atlantic}, which praised the gaming community's rigour, noting that ``gamers were better than scientists at catching fraud'' \cite{refDreamAtlantic}. 
This case underscores how even subtle, technically plausible manipulations -- such as tweaking PRNG behaviour -- can be weaponised, and how statistical forensics can be a crucial tool in identifying such deceit.

\vspace{0.5em}

While not exhaustive, the five cases presented in the material illustrate several common manipulation strategies. 
Other known fraudulent runs include a submission by \textit{Mekarazium}. 
This run involved spliced footage from multiple pre-recorded attempts, presented as a single live-streamed gameplay session during a \textit{Games Done Quick (GDQ)} event~\cite{refKJMekarazium, refAbyssoftMekarazium, refGDQMekaraziumReupload}. 
A heavily spliced \textit{Diablo} run was also reported--fifteen years after its submission--having been composed of multiple segments edited to appear as a single run. It was ultimately caught through an extremely thorough analysis by a group of experts of the game~\cite{refAbyDiablo, refDwangoDiablo}. 
Together, aforementioned incidents reveal a growing taxonomy of deception methods and underscore the need for community-supported forensic frameworks -- such as \textsc{Tracer} -- to help reviewers detect, categorise, and respond to such manipulations with greater consistency and discipline.

\bibliographystyle{unsrt}
\bibliography{Tracer_Authors_Version_new}

\end{document}